\documentstyle[psfig]{l-aa}
\begin{document}
\thesaurus{09.13.2, 11.09.4, 11.14.1, 11.19.3, 13.19.1}%
%
%==================================================================%%
%%                                                                  %%
%%                                                                  %%
%%                                                                  %%
%%                      A S T R O N O M Y                           %%
%%                                                                  %%
%%                           AND                                    %%
%%                                                                  %%
%%                  A S T R O P H Y S I C S                         %%
%%                                                                  %%
%%                                                                  %%
%%        LaTeX Support                             Version 2.09    %%
%%                                                                  %%
%%==================================================================%%
%
%  Abbreviations
%
\def\etal {et al.}
\def\ie {i.\,e.}
\def\etseq {{\em et seq.}}
\def\vs {{it vs.}}
\def\perse {{it per se}}
\def\adhoc {{\em ad hoc}}
\def\eg {e.\,g.}
\def\etc {etc.}
\def\ccpers {\hbox{${\rm cm}^3{\rm s}^{-1}$}}
\def\DEGR {\hbox{$^{\circ }$}}
\def\vlsr {\hbox{${v_{\rm LSR}}$}}
\def\vel {\hbox{${v_{\rm LSR}}$}}
\def\vhel {\hbox{${v_{\rm HEL}}$}}
\def\delv {\hbox{$\Delta v_{1/2}$}}
\def\dvel {\hbox{$\Delta v_{1/2}$}}
\def\TL {$T_{\rm L}$}
\def\TC {$T_{\rm c}$}
\def\TEX {$T_{\rm ex}$}
\def\TMB {$T_{\rm MB}$}
\def\TKIN {$T_{\rm kin}$}
\def\TREC {$T_{\rm rec}$}
\def\TSYS {$T_{\rm sys}$}
\def\TVIB {$T_{\rm vib}$}
\def\TROT {$T_{\rm rot}$}
\def\TDUST {$T_{\rm d}$}
\def\TASTAR {$T_{\rm A}^{*}$}
\def\TVIBST {$T_{\rm vib}^*$} 
\def\H0 {$H_{\rm o}$}
\def\mic {$\mu\hbox{m}$}
\def\micro {\mu\hbox{m}}
\def\SDOZ {\hbox{$S_{12\mu \rm m}$}}
\def\STWE {\hbox{$S_{25\mu \rm m}$}}
\def\SSIX {\hbox{$S_{60\mu \rm m}$}}
\def\SHUN {\hbox{$S_{100\mu \rm m}$}}
\def\solmass {\hbox{M$_{\odot}$}}
\def\solum {\hbox{L$_{\odot}$}}
\def\irlum {\hbox{$L_{\rm IR}$}}
\def\ohlum {\hbox{$L_{\rm OH}$}}
\def\blum {\hbox{$L_{\rm B}$}}
\def\numd {\hbox{$n\,({\rm H}_2$)}}                   
\def\rhounit {$\hbox{M}_\odot\,\hbox{pc}^{-3}$}
\def\kms {\hbox{${\rm km\,s}^{-1}$}}
\def\kmsyr {\hbox{${\rm km\,s}^{-1}\,{\rm yr}^{-1}$}}
\def\kmsmpc {\hbox{${\rm km\,s}^{-1}\,{\rm Mpc}^{-1}$}} 
\def\Kkms {\hbox{${\rm K\,km\,s}^{-1}$}}
\def\percc {$\hbox{{\rm cm}}^{-3}$}    %cm-3
\def\cmsq  {$\hbox{{\rm cm}}^{-2}$}    %cm-2
\def\cmsix  {$\hbox{{\rm cm}}^{-6}$}  %cm-6
\def\arcsec {\hbox{$^{\prime\prime}$}}
\def\arcmin {\hbox{$^{\prime}$}}
\def\ffam {\hbox{$\,.\!\!^{\prime}$}}
\def\ffas {\hbox{$\,.\!\!^{\prime\prime}$}}
\def\ffM {\hbox{$\,.\!\!\!^{\rm M}$}}
\def\ffm {\hbox{$\,.\!\!\!^{\rm m}$}}
\def\HI  {\hbox{HI}}
\def\HII {\hbox{HII}}
%
%   Greek and abbreviations for radio recomb lines etc
%
\def \AL {$\alpha $}    % gr. alpha
\def \BE {$\beta $}     % gr. beta
\def \GA {$\gamma $}    % gr. gamma
\def \DE {$\delta $}    % gr. delta
\def \EP {$\epsilon $}  % gr. epsilon
\def \alde {($\Delta \alpha ,\Delta \delta $)}
\def \MU {$\mu $}       % gr. mue
\def \TAU {$\tau $}     % gr. tau
\def \tapp {$\tau _{\rm app}$}
\def \tuns {$\tau _{\rm uns}$}
\def \RH {\hbox{$R_{\rm H}$}}         % OH main line ratio
\def \RT {\hbox{$R_{\rm \tau}$}}      % OH main tau  ratio
\def \BN  {\hbox{$b_{\rm n}$}}        % bn
\def \BETAN {\hbox{$\beta _n$}}       % beta factor
\def \TE {\hbox{$T_{\rm e}$}}         % Electron Temp.
\def \NE {\hbox{$N_{\rm e}$}}         % Electron Dens.
% molecules
%
\def\MOLH {\hbox{${\rm H}_2$}}                    %H2
\def\HDO {\hbox{${\rm HDO}$}}                     %HDO
\def\AMM {\hbox{${\rm NH}_{3}$}}                  %NH3
\def\NHTWD {\hbox{${\rm NH}_2{\rm D}$}}           %NH2D
\def\CTWH {\hbox{${\rm C_{2}H}$}}                 %C2H
\def\TCO {\hbox{${\rm ^{12}CO}$}}                 %12CO
\def\CEIO {\hbox{${\rm C}^{18}{\rm O}$}}          %C18O
\def\CSEO {\hbox{${\rm C}^{17}{\rm O}$}}          %C17O
\def\CTHFOS {\hbox{${\rm C}^{34}{\rm S}$}}        %C34S
\def\THCO {\hbox{$^{13}{\rm CO}$}}                %13CO
\def\WAT {\hbox{${\rm H}_2{\rm O}$}}              %H2O
\def\WATEI {\hbox{${\rm H}_2^{18}{\rm O}$}}       %H218O
\def\CYAN {\hbox{${\rm HC}_3{\rm N}$}}            %HC3N
\def\CYACFI {\hbox{${\rm HC}_5{\rm N}$}}          %HC5N
\def\CYACSE {\hbox{${\rm HC}_7{\rm N}$}}          %HC7N
\def\CYACNI {\hbox{${\rm HC}_9{\rm N}$}}          %HC9N
\def\METH {\hbox{${\rm CH}_3{\rm OH}$}}           %CH3OH
\def\MECN {\hbox{${\rm CH}_3{\rm CN}$}}           %CH3CN
\def\METAC {\hbox{${\rm CH}_3{\rm C}_2{\rm H}$}}  %CH3C2H
\def\CH3C2H {\hbox{${\rm CH}_3{\rm C}_2{\rm H}$}} %CH3C2H
\def\FORM {\hbox{${\rm H}_2{\rm CO}$}}            %H2CO
\def\MEFORM {\hbox{${\rm HCOOCH}_3$}}             %HCOOCH3
\def\THFO {\hbox{${\rm H}_2{\rm CS}$}}            %H2CS
\def\ETHAL {\hbox{${\rm C}_2{\rm H}_5{\rm OH}$}}  %C2H5OH
\def\CHTHOD {\hbox{${\rm CH}_3{\rm OD}$}}         %CH3OD
\def\CHTDOH {\hbox{${\rm CH}_2{\rm DOH}$}}        %CH2DOH
\def\CYCP {\hbox{${\rm C}_3{\rm H}_2$}}           %C3H2
\def\CTHHD {\hbox{${\rm C}_3{\rm HD}$}}           %C3HD
\def\HTCN {\hbox{${\rm H^{13}CN}$}}               %H13CN
\def\HNTC {\hbox{${\rm HN^{13}C}$}}               %HN13C
\def\HCOP {\hbox{${\rm HCO}^+$}}                  %HCO+
\def\HTCOP {\hbox{${\rm H^{13}CO}^{+}$}}          %H13CO+
\def\NNHP {\hbox{${\rm N}_2{\rm H}^+$}}           %N2H+
\def\CHTHP {\hbox{${\rm CH}_3^+$}}                %CH3+
\def\CHP {\hbox{${\rm CH}^{+}$}}                  %CH+
\def\ETHCN {\hbox{${\rm C}_2{\rm H}_5{\rm CN}$}}  %C2H5CN
\def\DCOP {\hbox{${\rm DCO}^+$}}                  %DCO+
\def\HTHP {\hbox{${\rm H}_{3}^{+}$}}              %H3+ 
\def\HTWDP {\hbox{${\rm H}_{2}{\rm D}^{+}$}}      %H2D+
\def\CHTWDP {\hbox{${\rm CH}_{2}{\rm D}^{+}$}}    %CH2D+
\def\CNCHPL {\hbox{${\rm CNCH}^{+}$}}             %CNCH+
\def\CNCNPL {\hbox{${\rm CNCN}^{+}$}}             %CNCN+
%
% Abbreviations for T. Wiklind article
%
\def\In {\hbox{$I^{n}(x_{\rm k},y_{\rm k},u_{\rm l}$})}
\def\Iobs {\hbox{$I_{\rm obs}(x_{\rm k},y_{\rm k},u_{\rm l})$}}
\def\Ingl {I^{n}(x_{\rm k},y_{\rm k},u_{\rm l})}
\def\Iobsgl {I_{\rm obs}(x_{\rm k},y_{\rm k},u_{\rm l})}
\def\Pbgl {P_{\rm b}(x_{\rm k},y_{\rm k}|\zeta _{\rm i},\eta _{\rm j})}
\def\Pbgm {P(x_{\rm k},y_{\rm k}|r_{\rm i},u_{\rm l})}
\def\Pbgn {P(x,y|r,u)}
\def\Pugm {P_{\rm u}(u_{\rm l}|w_{\rm ij})}
\def\Pdem {P_{\rm b}(x,y|\zeta (r,\theta ),\eta (r,\theta ))} 
\def\Pden {P_{\rm u}(u,w(r,\theta ))}
\def\greekgl {(\zeta _{\rm i},\eta _{\rm j},u_{\rm l})}
\def\greekg1 {(\zeta _{\rm i},\eta _{\rm j})}
\title{Dense gas in nearby galaxies}
\subtitle{XI. H$_{2}$CO and CH$_{3}$OH: Molecular abundances and 
physical conditions}
\author{S.~H\"{u}ttemeister\inst{1,2}, R.~Mauersberger\inst{3,4},
C.~Henkel\inst{3} }
\offprints{S. H{\"u}ttemeister, RAIUB}
\institute{
 Radioastronomisches Institut der Universit\"{a}t Bonn,
 Auf dem H\"{u}gel 71, D - 53121 Bonn, Germany
\and
 Harvard-Smithsonian Center for Astrophysics,
 60 Garden Street, Cambridge, MA 02138, U.S.A
\and
 Max-Planck-Institut f{\"u}r Radioastronomie,
 Auf dem H{\"u}gel 69, D - 53121 Bonn, Germany
\and
 Steward Observatory, The University of Arizona,
 Tucson, AZ 85721, U.S.A.}
\date{received 7 November 1996; accepted 29 April 1997 }
\maketitle
\begin{abstract} Multilevel observations of formaldehyde (\FORM ) 
and methanol (\METH ) toward the nearby spiral galaxies
NGC\,253, Maffei\,2, IC342, M\,82 and NGC\,6946 are presented.
\FORM\ was detected in all galaxies (tentatively in NGC\,6946). \METH\
was detected in all objects with the notable exception of M\,82.

H$_2$CO line intensity ratios point out differences in gas density both
between galaxies and within the central regions of individual objects.
Model calculations show that the bulk of the gas emitting \FORM\ in
NGC\,253 is at a density of $\sim 10^4$\,\percc , while the
\FORM\ lines in M\,82 and IC\,342 trace two different, spatially
separated gas components with densities of $\leq 10^4$\,\percc\ and
$\sim 10^{6}$\,\percc . The south-western molecular hotspot in M\,82
and the center of IC\,342 are the regions with the highest density.

Methanol is subthermally excited in all galaxies, with the lowest
excitation temperatures found in IC\,342. The \METH\ abundance in
NGC\,253 and the non-starburst nuclei of IC\,342 and Maffei\,2 are
comparable. A map of the $3_{\rm k}-2_{\rm k}$ lines in NGC\,253 shows
that \METH\ traces clumpy structures better than other molecules
requiring high gas density to be excited.  \METH\ toward M\,82 is at
least an order of magnitude less abundant than in otherwise comparable
galaxies. This confirms the existence of global chemical differences,
and thus very large scale variations in the state of the molecular gas
phase, even between galaxies commonly classified as starburst nuclei.
\end{abstract} 
\section{Introduction}
The central regions of galaxies, including our own, are characterized
by a unique population of molecular clouds, both denser and warmer than
typical disk clouds. In galaxies undergoing a nuclear starburst, these
clouds are subject to turbulence and intense UV radiation, as well as
shocks, due to both a steep gravitational potential and the formation
and death of young massive stars.  They form a distinct type of
interstellar molecular environment, different from Galactic dark clouds
and the surroundings of H\,{\sc II} regions. Galactic center clouds are a
rich source of molecular emission; $\sim$\,25 molecular species have
been discovered in the nuclear regions of external galaxies.

Formaldehyde and methanol are both useful tracers of the dense
interstellar gas phase.  Their molecular structure is more complicated
than that of HCN or CS, the ``classical'' tracers of dense gas.  This
leads to a greater wealth of transitions which can be observed in the
millimeter wavelength range, providing a powerful tool to understand
the physical and chemical conditions of the gas associated with
starbursts.
 
The first {\em formaldehyde} transitions detected in external galaxies
were $K$-doublet ortho transitions in the centimeter wavelength range 
($1_{10}-1_{11}$ at 6\,cm and $2_{11}-2_{12}$ at 2\,cm; Gardner
\& Whiteoak 1974, 1976, 1979, Whiteoak \& Gardner 1976, Cohen
\etal\ 1979, Graham \etal\ 1978, Seaquist \& Bell 1990, Baan \& Goss
1992).  Galaxies where these lines have been detected include NGC\,253,
NGC\,4945, M\,82, Centaurus\,A, NGC\,3628, M\,31 and the LMC. Recently,
the $2_{11}-2_{12}$ transition has been detected toward the
Einstein Ring gravitational lens system B0218+357 at a redshift of
$z=0.68$ (Menten \& Reid 1996).  In all the above sources the lines are
observed in absorption; with \TEX $<$ 2.7\,K, they can absorb even the
microwave background.  The $1_{10}-1_{11}$ line emission
was first observed toward Arp\,220 (Baan \etal\ 1986). Detections in
nine other galaxies and an interferometric map of Arp\,220 (Baan \etal\
1993, Baan \& Haschick 1995) followed. These lines are interpreted as
\FORM\ megamasers.

There are less observations of the rotational transitions at millimeter
wavelengths, even though these transitions are sensitive tracers of
cloud temperature and density. The $3_{03}-2_{02}$ para transition
was discovered in M\,82 (Baan \etal\ 1990), the $3_{21}-2_{20}$ para
line was observed in NGC\,253 (Petuchowski \& Bennett 1992), and the
$3_{12}-2_{11}$ ortho transition was detected in the LMC by Johansson
\etal \ (1994) and tentatively in the southern galaxy NGC\,4945
(Mauersberger et al.\ 1996b).

The first successful observation of extragalactic {\em methanol} (\METH
) was reported in 1987 by Henkel \etal . They detected a superposition
of four $2_{\rm k}-1_{\rm k}$ transitions at 96\,GHz in NGC\,253,
IC\,342 and (tentatively) NGC\,6946.  This group of transitions
requires only low excitation and all four components are among the
strongest methanol lines observed in Galactic sources. The same group
of transitions was detected in the southern starburst galaxy NGC\,4945
(Henkel \etal\ 1990). The $0_0-1_{-1}E$ transition was seen toward
NGC\,253 by Henkel \etal \ (1993).  Methanol masers, very common in the
Galaxy, are also detected toward the Large Magellanic Cloud at 6.6\,GHz
(Sinclair \etal\ 1992, Ellingsen \etal\ 1994).  Here, we report
observations of six \FORM\ and seven \METH\ rotational transitions
toward five galaxies, thus considerably expanding the existing
database.

\section{Observations}

\begin{table}
\begin{flushleft}
\caption{\label{log} Summary of the observed transitions. For \FORM ,
$o$ and $p$ denote ortho and para transitions.}
\begin{tabular}{lrrr}
\multicolumn{1}{c}{Line} & \multicolumn{1}{c}{Frequency} &
\multicolumn{1}{c}{E$_{\rm u}^{\rm a)}$} &
\multicolumn{1}{c}{$\theta$} \\
 & \multicolumn{1}{c}{GHz} & \multicolumn{1}{c}{K} 
 & \multicolumn{1}{c}{$''$} \\
\hline
\multicolumn{4}{l}{ \FORM } \\
$2_{12}-1_{11}\ o$ & 140.840 &  17   & 16 \\
$2_{02}-1_{01}\ p$ & 145.603 &  10   & 16 \\
$2_{11}-1_{10}\ o$ & 150.498 &  23   & 16 \\
$3_{13}-2_{12}\ o$ & 211.211 &  32   & 12 \\
$3_{03}-2_{02}\ p$ & 218.222 &  21   & 12 \\
$3_{12}-2_{11}\ o$ & 225.698 &  33   & 12 \\
\hline 
\multicolumn{4}{l}{ \METH } \\
$5_{-1}-4_{0}$E            & 84.521   & 32      & 27 \\
$2_{\rm k}-1_{\rm k}^{b)}$ & 96.741   & 5--20   & 27 \\
$3_{1}-2_{1}$A$^{+}$       & 143.866  & 28      & 16 \\
$3_{\rm k}-2_{\rm k}^{c)}$ & 145.103  & 12--51  & 16 \\
$7_{1}-8_{0}$E             & 220.07   &  88     & 12 \\
$8_{-1}-7_{0}$E            & 229.75   & 81      & 12 \\
$3_{-2}-4_{-1}$E           & 230.027  & 32      & 12 \\
\hline \\
\end{tabular} \\
$a)$ Energy of the upper level; for E-type methanol, the values refer
to the $1_1$-level, 6.9\,K above ground. \\
$b)$ Lines contributing to the $2_{\rm k}-1_{\rm k}$ group:
$2_{-1}-1_{-1}E$ (5\,K), $2_0-1_0A^+$ (6\,K), $2_0-1_0E$ (12\,K),
$2_1-1_1E$ (20\,K) \\ 
$c)$ Lines contributing to the $3_{\rm k}-2_{\rm k}$ group:
$3_{0}-2_{0}E$ (19\,K), $3_{-1}-2_{-1}E$ (12\,K),
$3_{0}-2_{0}A^{+}$ (14\,K), $3_{2}-2_{2}A^{-}$ (51\,K),
$3_{2}-2_{2}E$ (28\,K), $3_{-2}-2_{-2}E$ (32\,K), $3_{1}-2_{1}E$
(27\,K), $3_{2}-2_{2}A^{+}$ (51\,K)
\end{flushleft}
\end{table}

All observations were carried out with the IRAM 30-m antenna.  The
measurements were made with SIS receivers tuned to a single sideband
with image sideband rejections of typically 7\,dB. An error in the 
determination of the sideband rejection of $\pm$1\,dB would result in
an additional calibration uncertainty of $\pm$5\%.  The temperature
scale used throughout this paper is main-beam brightness temperature
(\TMB ).  System temperatures were between 540\,K and 580\,K \TMB\ in
the 3\,mm wavelength range, between 600\,K and 900\, \TMB\ at 2\,mm
wavelength and between 1200\,K and 1800\,K \TMB\ at 1.3\,mm. All
transitions were observed with either a 512 $\times$ 1\,MHz filterbank
or a 864 channel acousto-optical spectrometer (AOS) with a total 
bandwidth of 500\,MHz.

The observed transitions, including their designations, frequencies and
energies above ground (Mangum \& Wootten 1993, Anderson \etal\ 1990),
and the beamwidth of the telescope, are summarized in
Table\,\ref{log}.

For broad lines encountered in external galaxies, it is not only
the system noise that determines the accuracy of line intensies and 
ratios, but also the flatness of the spectrometer bandbass, which is
affected by the stability of the weather and the electronic components. 
We used a wobbling secondary mirror, switching with a frequency of 
0.5\,Hz between the source and two reference positions placed 
symmetrically at 4\arcmin\ offset in azimuth. The backends are of 
high stability. Thus, the baselines we obtained are very flat. 
Therefore, only baselines of order zero or one were subtracted from 
the spectra, and the data were smoothed to a resolution of 
$\sim$\,10\,\kms.

From continuum cross scans on nearby sources, made every $\sim$
\,3\,hours, we estimate the pointing accuracy to be of order
5\arcsec\ or better. To check the calibration of the telescope, we
usually observed line emission from strong Galactic sources like W3(OH)
in each period to achieve internal consistency.  We expect the
calibration to be correct on a level of $\sim \pm 15$\%.

\section{Results}

\subsection{Formaldehyde}
\begin{figure*}
\psfig{file=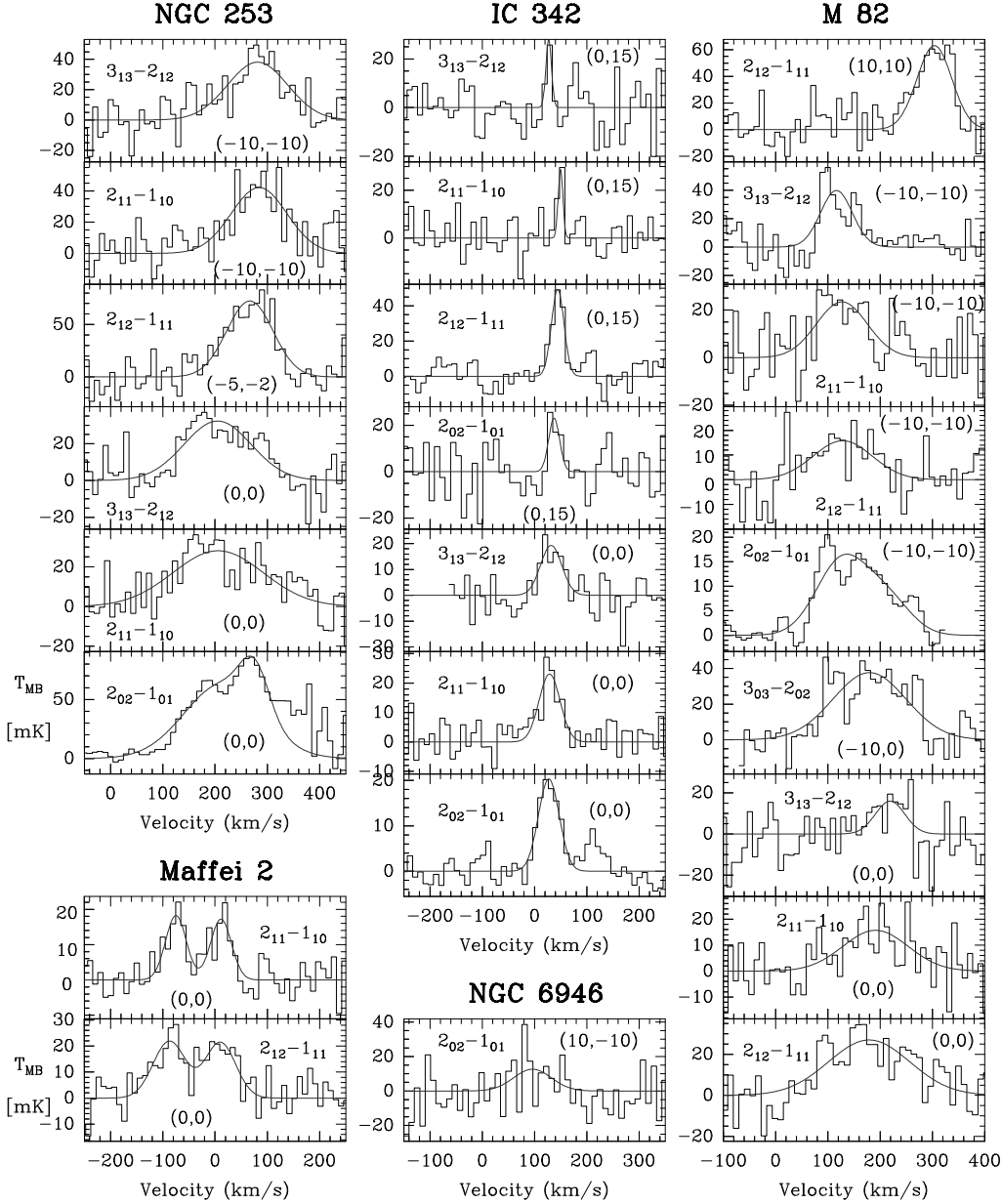,bbllx=0pt,bblly=80pt,bburx=555pt,bbury=760pt,height=17.0cm}
\caption{\label{formf1} H$_2$CO spectra for those positions and
transitions where we detected, at least tentatively, a line. All
spectra have been smoothed to a resolution of 10\,\kms. The central
positions (1950.0) and distances of the galaxies are:  NGC\,253:
$\alpha = 00^{\rm h}45^{\rm m}06.0^{\rm s}$, $\delta =
-25^{\circ}33'36''$, 2.5\,Mpc, Maffei\,2: $\alpha = 02^{\rm h}38^{\rm
m}08.5^{\rm s}$, $\delta = 59^{\circ}23'24''$, 5\,Mpc, IC\,342: $\alpha
= 03^{\rm h}41^{\rm m}57.5^{\rm s}$, $\delta = 67^{\circ}56'25''$,
1.8\,Mpc, M\,82: $\alpha = 09^{\rm h}51^{\rm m}43.0^{\rm s}$, $\delta =
69^{\circ}55'00''$, 3.3\,Mpc, NGC\,6946:  $\alpha = 20^{\rm h}33^{\rm
m}48.0^{\rm sec}$, $\delta = 59^{\circ}59'00'$, 11\,Mpc.}
\end{figure*}

The parameters of Gaussian fits to the detected lines are given
in Table\,\ref{formt1}. Spectra are displayed in
Fig.\ \ref{formf1}. Since not all lineshapes are exactly Gaussian, we
have also determined the integrated line intensities
by adding the channels with emission, calculating the errors from the 
rms noise per channel. The results are identical, with the errors of the
Gaussian typically being slightly larger than the errors obtained using the 
rms noise. 

\begin{table}
\begin{flushleft}
\caption{\label{formt1} Parameters of \FORM\ lines observed 
toward nearby spiral galaxies. The offsets from the central position are
given in arcseconds. Errors (1$\sigma$) are given in parenthesis. }
\vspace{0.5mm}
\begin{tabular}{l@{\hspace{1mm}}l|r@{\hspace{1mm}}r@{\hspace{1mm}}r@{\hspace{1mm}}r}
\small
Source & Trans.  & 
\multicolumn{1}{c}{$\int T_{\rm MB} dv$} &
\multicolumn{1}{c}{\vlsr} &
\multicolumn{1}{c}{$\Delta v_{1/2}$} &
\multicolumn{1}{c}{\TMB} \\
    &  &  \multicolumn{1}{c}{\Kkms }  
 & \multicolumn{2}{c}{\kms } 
& \multicolumn{1}{c}{mK}  \\
\hline
NGC253  &  & &  & \\
(0,0)  & $2_{02}-1_{01}^{\rm a),b)}$  & $\sim 10$ & {\it 210} & {\it 170} 
            &  $\sim 45$ \\
       & $3_{03}-2_{02}$   &  $\leq 2.8^{\rm c)}$  & --   & -- 
            & $\leq 17^{\rm d)}$ \\
       & $2_{11}-1_{10}$  & 6.1(0.7)  & 206(7)  & 204(25) & 28(10) \\
       & $3_{13}-2_{12}$  & 5.2(0.7)  & 205(10) & 151(20) & 32(11) \\
(--5,--2)  & $2_{12}-1_{11}$ &  8.0(0.9)  & 267(6)  & 104(13) & 72(13) \\
(--10,--10)  & $2_{11}-1_{10}$   & 5.6(0.8)  & 284(7)  & 126(23) & 42(11) \\
            & $3_{13}-2_{12}$  & 5.5(0.7)  & 281(8)  & 136(26) & 38(11) \\
\hline
Maffei2  &  & &  & \\
(0,0)     & $3_{03}-2_{02}$  & $\leq 4.9$ & --  & --  & $\leq$33  \\
          & $2_{12}-1_{11}$   & 1.7(0.4)  & --87(6) & 74(18)  & 22(6) \\
          &               & 1.6(0.3)  & 10(6)   & 69(17)  & 21(6) \\
          & $2_{11}-1_{10}$ & 0.9(0.2)  & --75(4) & 47(15)  & 19(4) \\
          &               & 0.9(0.2)  & 12(5)   & 45(11)  & 18(4) \\
          & $3_{13}-2_{12}$  &  $\leq 1.8$  & --  & --  & $\leq$12  \\
          & $3_{12}-2_{11}$  & $\leq 1.6$ & --  & --  & $\leq$11 \\
\hline
IC342   &  & &  & \\
(0,0)    &  $2_{02}-1_{01}^{\rm b)}$   & 1.0(0.1) & 27(2)  & 47(5)  & 20(3)  \\
         &  $2_{12}-1_{11}$   & $\leq 0.9$  & --  & --  & $\leq$16  \\
         &  $2_{11}-1_{10}$   & 1.2(0.2)  & 29(4)   & 47(11)  & 23(5) \\
         &  $3_{13}-2_{12}$  & 1.0(0.2)  & 32(6)    & 48(12)  & 19(7) \\
(0,5)    &  $3_{03}-2_{02}$ & $\leq 3.3$  & -- & --  &  $\leq$40 \\
(0,15)   &  $2_{02}-1_{01}$  & 0.6(0.2)  & 38(5)   & 25(5)   & 23(10) \\
         &  $2_{12}-1_{11}$  & 1.5(0.2)  & 44(2)   & 28(5)   & 49(7) \\
         &  $2_{11}-1_{10}$   & 0.3(0.2)$^{\rm e)}$ & 50(2) & 11(3) & 29(7) \\
         &  $3_{13}-2_{12}$   & 0.4(0.2)$^{\rm e)}$  & 27(3) & 13(6) & 28(9) \\
         &  $3_{12}-2_{11}$  & $\leq 0.9$ & --  & -- &  $\leq$12 \\
\hline
M82    &  & &  & \\
(0,0)   & $2_{02}-1_{01}^{\rm a)}$ & $\leq 3.0$  & --  & -- &  $\leq$30 \\
        & $2_{12}-1_{11}$ & 5.3(0.9)  & 179(17) & 183(37) &  27(9) \\
        & $2_{11}-1_{10}$ & 2.5(0.6)  & 191(18) & 147(44)  & 16(9) \\
        & $3_{13}-2_{12}$  & 1.3(0.4)$^{\rm e)}$ & 219(16) & 66(25) & 16(10) \\
(--10,0)    & $3_{03}-2_{02}^{\rm f)}$  &  6.6(0.5) & 180(6) & 165(12) & 38(8)\\
(--10,--10) & $2_{02}-1_{01}^{\rm a),b)}$ & $\sim 1.8$  & {\it 120} 
            & {\it 110} &  $\sim 15$ \\
            & $2_{12}-1_{11}$  & 2.2(0.5) & 128(18)  & 133(31) &  16(10) \\
            &  $2_{11}-1_{10}$  & 2.9(0.6)  & 127(10) & 114(34)  & 23(11) \\
            &   $3_{13}-2_{12}$ & 3.0(0.5)  & 117(8)  & 71(14)   & 40(11) \\
(10,10)  & $2_{12}-1_{11}$ & 5.3(0.6) & 304(5) & 79(10) & 63(13) \\
         &  $2_{11}-1_{10}$ & $\leq 0.8$  & --  & -- &  $\leq$10 \\
\hline
NGC6946 &  & &  & \\
(10,-10)  & $2_{02}-1_{01}$ & 1.1(0.5)$^{\rm e)}$ & 96(18) & 85(22) & 13(11) \\
\hline
\end{tabular} \\
\footnotesize
a) In NGC\,253 and M\,82, the $2_{0,2}-1_{0,1}$-line is blended with the 
16--15 transition of HC$_3$N, offset by 42\,MHz (86.5\,\kms). An approximate 
fit for NGC\,253 was obtained by fixing the central velocity and width of 
the HC$_3$N line to the values known from the 15--14 and 17--16 HC$_3$N 
transitions (Mauersberger et al.\ 1990). For M\,82, the \FORM\ line
parameters were fixed to the mean value derived from unblended \FORM\ 
lines in the same position, assuming the nominal offset for HC$_3$N. 
Both estimates are indicated by italics. \\
b) Data from Mauersberger \etal\ 1995 \\
c) Limits for the integrated intensity are 3$\sigma$ limits, based on the 
rms per channel extrapolated to a linewidth fixed to the value determined
from detected \FORM\ transitions toward the same position. \\
d) All limits and errors to \TMB\ are rms in a 10\,\kms\ wide channel \\
e) tentative \\
f) Data from Baan \etal\ 1990 \\
\end{flushleft}
\normalsize
\end{table}

\normalsize

We detected mm wavelength \FORM\ transitions toward all galaxies that
were observed, although the detection of the $2_{02}-1_{01}$ line
in NGC\,6946 is tentative.

Toward NGC\,253, four of the five transitions observed could be
detected; in M\,82, all five lines were found. The three positions
observed in M\,82 are detected in the $2_{12}-1_{11}$ line, while the
$2_{11}-1_{10}$ line is not seen toward the (10\arcsec ,10\arcsec)
offset. Toward this position, the $J=2-1$ twin lines thus have
different intensities.

In IC\,342, four out of six transitions were detected.  It is
interesting to note that the $3_{13}-2_{12}$ transition was detected
both toward the central and (tentatively) toward an offset position, 
while its twin transition, the $3_{12}-2_{11}$ line, only observed 
toward the offset position, was not detected at an rms level of 
$\sim$\,12\,mK. As in M\,82, the intensities of the $J=2-1$ twin 
lines differ. Toward the center, the $2_{11}-1_{10}$ line is 
considerably stronger than its counterpart, the $2_{12}-1_{11}$ 
transition, which is not detected. Toward a northern offset position 
however, the $2_{12}-1_{11}$ line is stronger than the (tentatively 
detected) $2_{11}-1_{10}$ line.

Toward Maffei\,2, two out of four lines were found. The profiles show a
double peaked structure. This double feature is also visible in the
HNC(1--0) transition (H\"uttemeister \etal\ 1995), but is not seen in
other molecular species.

\subsection{Methanol}
\begin{figure}
\psfig{file=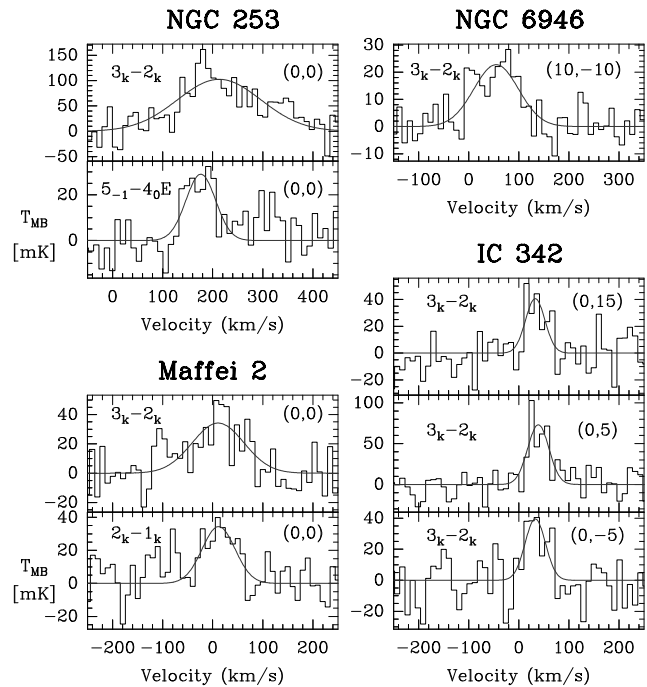,bbllx=65pt,bblly=300pt,bburx=555pt,bbury=675pt,height=9.8cm}
\caption{\label{methf1} \METH\ transitions detected toward NGC\,253,
Maffei\,2, IC\,342 and NGC\,6964. }
\end{figure}

\begin{figure*}
\psfig{file=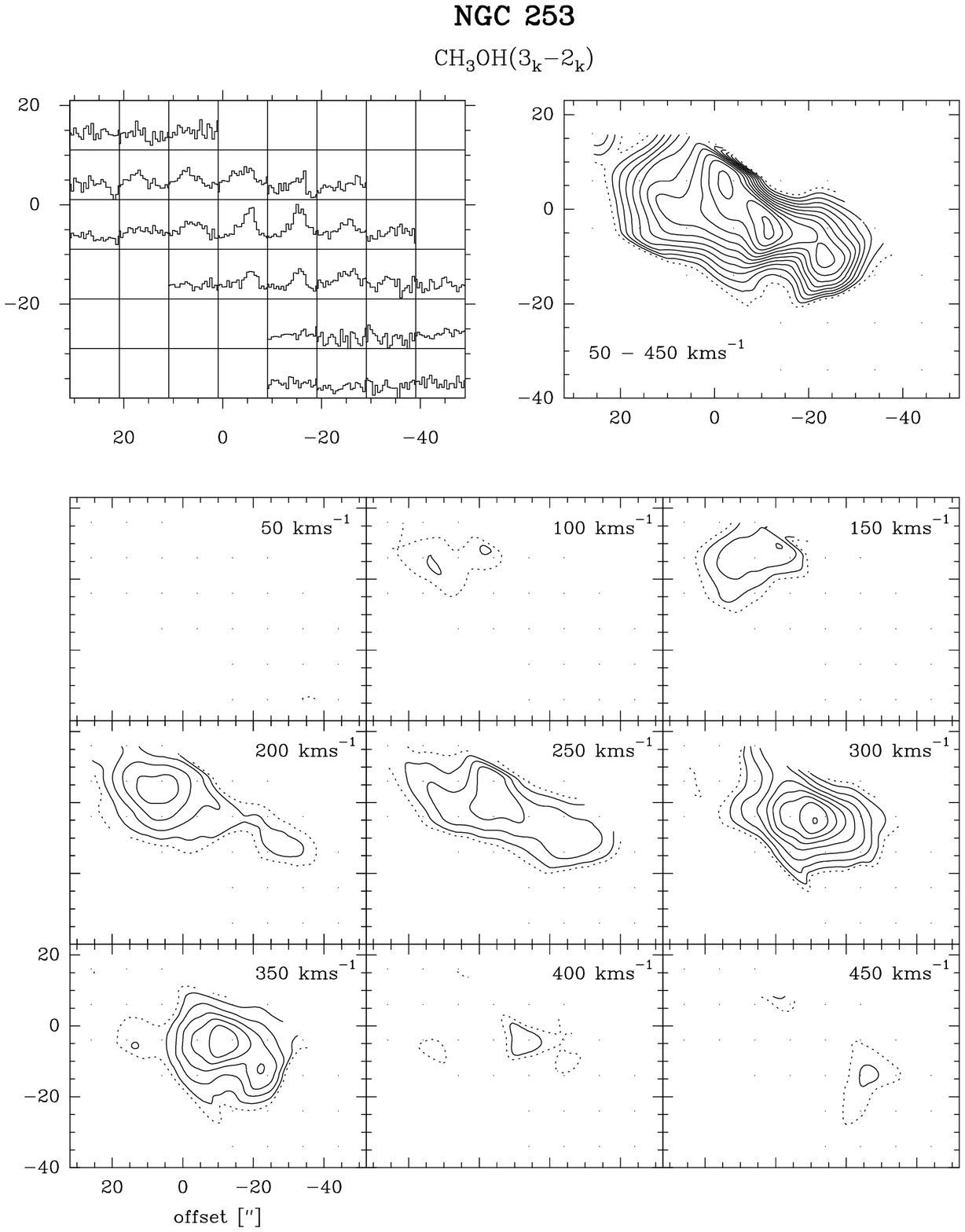,bbllx=0pt,bblly=80pt,bburx=555pt,bbury=700pt,height=16.7cm}
\caption{\label{methf2} Map of \METH\ ($3_{\rm k}-2_{\rm k}$) toward 
NGC\,253. The upper left panel shows the spectra with a \vlsr\ range of 50--450\,kms\ and \TMB\ ranging from --70 to 200 mK. The total integrated
intensity map is given in the upper left panel; solid contours start at
5.25\,K\,\kms\ (3$\sigma$) in steps of 1.75\,K\,\kms\ (1$\sigma$). The dashed 
contour denotes the 2$\sigma$ level at 3.5\,K\,\kms. In the lower panel, 
we present channel maps of 50\,\kms\ width. The first solid contour is at
1.8\,K\,\kms\ (3$\sigma$), the steps are 0.9\,K\,\kms; the dashed contour is 
1.2\,K\,\kms\ (2$\sigma$). }
\end{figure*}

The results of  Gaussian fits for all observations of single positions 
are given in Table\,\ref{metht1}; spectra are displayed in Fig.\ 
\ref{methf1}. We also obtained a map of NGC\,253 in the $J_{\rm k} = 
3_{\rm k}-2_{\rm k}$ transitions at 145\,GHz with a 10$''$ spacing
(Fig.\,\ref{methf2}).

\begin{table}
\begin{flushleft}
\caption{\label{metht1} Parameters of \METH\ lines observed toward 
nearby spiral galaxies. The offsets from the central position are
given in arcseconds.}
\vspace{2mm}
\begin{tabular}{l@{\hspace{1mm}}l|r@{\hspace{1mm}}r@{\hspace{1mm}}r@{\hspace{1mm}}r}
\small
Source & Trans. & 
\multicolumn{1}{c}{$\int T_{\rm MB} dv$} &
\multicolumn{1}{c}{\vlsr} &
\multicolumn{1}{c}{$\Delta v_{1/2}$} &
\multicolumn{1}{c}{\TMB} \\
  & & 
  \multicolumn{1}{c}{\Kkms }  
& \multicolumn{2}{c}{\kms }  
& \multicolumn{1}{c}{mK}  \\
\hline
NGC253 & & & & & \\
(0,0)     & $5_{-1}-4_{0}E$     & 2.2(0.4)  & 176(6)  &  70(12) & 29(9) \\
          & $2_{\rm k}-1_{\rm k}^{\rm a)}$ & 3.8(0.6) & 190(5) 
              & 70(12)  &  51(15) \\
          &   &  14.9(0.7) & 309(2) & 104(6) & 136(15) \\
          & $3_{1}-2_{1}A^{+}$  & $\leq 2.7^{\rm b)}$ & -- & -- 
              & $\leq 20(18)$ \\
           & $3_{\rm k}-2_{\rm k}$ & 21.4(1.8) & 213(8) 
              & 196(21) & 102(25) \\
           & $8_{-1}-7_{0}E$     & $\leq 2.0$ & -- & -- 
              & $\leq 15$ \\ 
           & $3_{-2}-4_{-1}E$    & $\leq 5.4$ & -- & -- & $\leq$19 \\
\hline
Maffei2 & & & & & \\
(0,0)     & $5_{-1}-4_{0}E$       & $\leq 1.6$ & -- & -- & $\leq$12 \\
          & $2_{\rm k}-1_{\rm k}$ & 2.8(0.7) & 12(10) 
              & 77(33)  & 34(12) \\
          & $3_{1}-2_{1}A^{+}$    & $\leq 0.6$ & -- & -- & $\leq$6 \\
          & $3_{\rm k}-2_{\rm k}$ & 4.4(0.7) & 10(10)  
              &  122(26) & 34(11) \\
          & $8_{0}-7_{1}E$        & $\leq 3.9$ & -- & -- & $\leq$39 \\
          & $8_{-1}-7_{0}E$       & $\leq 2.7$ & -- & -- & $\leq$27 \\
\hline
IC342   & & & & & \\
(0,0)   & $5_{-1}-4_{0}E$   & $\leq 1.1$ & -- & -- & $\leq$16 \\
        &  $3_{-2}-4_{-1}E$    & $\leq 2.1$ & -- & -- & $\leq$31 \\
(0,--5)   &  $3_{\rm k}-2_{\rm k}$ & 2.0(0.4) & 33(6) & 48(9) & 40(13) \\
(0,5)     &  $2_{\rm k}-1_{\rm k}^{\rm a)}$ & 4.6(0.3) &
                46(2) & 63(3) & 68(15) \\
          & $3_{\rm k}-2_{\rm k}$ & 3.9(0.6) & 39(3) & 50(9)  & 73(14) \\
(0,15)    & $5_{-1}-4_{0}E$       & $\leq 0.7$ & -- & -- & $\leq$11 \\
          & $3_{\rm k}-2_{\rm k}$ & 2.0(0.5) & 33(6) & 
               46(12) & 41(13) \\
          & $8_{0}-7_{1}E$        & $\leq 2.6$ & -- & -- & $\leq$39 \\
          & $3_{-2}-4_{-1}E$      & $\leq 1.5$ & -- & -- & $\leq$22 \\
\hline
M82   & & & & & \\
(0,0)     & $5_{-1}-4_{0}E$       & $\leq 1.1$ & -- & -- & $\leq$10 \\
          & $3_{\rm k}-2_{\rm k}$ & $\leq 2.3$ & -- & -- & $\leq$15 \\
          & $8_{0}-7_{1}E$        & $\leq 2.4$ & -- & -- & $\leq$18 \\
          & $3_{-2}-4_{-1}E$      & $\leq 1.6$ & -- & -- & $\leq$11 \\
(--10,0)  & $2_{\rm k}-1_{\rm k}^{\rm a)}$ & $\leq 3.3$ & -- & -- & $\leq$34 \\
(--10,--10) &   $5_{-1}-4_{0}E$   & $\leq 1.1$ & -- & -- & $\leq$10 \\ 
            &   $3_{\rm k}-2_{\rm k}$  & $\leq 1.8$ & -- & -- & $\leq$11 \\
            & $8_{0}-7_{1}E$  & $\leq 1.2$ & -- & -- & $\leq$11 \\
            & $3_{-2}-4_{-1}E$  & $\leq 1.8$ & -- & -- & $\leq$16 \\
(10,10)     &    $3_{\rm k}-2_{\rm k}$  & $\leq 1.7$ & -- & -- & $\leq$15 \\
            &  $8_{0}-7_{1}E$   & $\leq 2.5$ & -- & -- & $\leq$25 \\
\hline
NGC\,6946  & & & & & \\
(4,0)      & $2_{\rm k}-1_{\rm k}^{\rm a)}$ & 1.7(?) & 55(?) & 
               100(?)  &  17(?) \\
(10,--10)  & $5_{-1}-4_{0}E$ & $\leq 0.7$ & -- & -- &$\leq 6$ \\
           & $3_{1}-2_{1}A^{+}$    & $\leq 0.8$ & -- & -- & $\leq$7  \\
           & $3_{\rm k}-2_{\rm k}$ & 2.5(0.3) 
                & 54(7) & 105(15) & 23(5) \\
           & $8_{0}-7_{1}E$        & $\leq 1.5$ & -- & -- & $\leq$13 \\
           & $3_{-2}-4_{-1}E$      & $\leq 1.4$ & -- & -- & $\leq$12 \\
\hline \\
\end{tabular} \\
\footnotesize
a) Results taken from Henkel \etal\ 1987. \\
b) Limits to the integrated intensity and \TMB\ were derived as for \FORM\
(Table \ref{formt1}) \\
\end{flushleft}
\normalsize
\end{table}

As is the case for the $J=2_{\rm k}-1_{\rm k}$ transitions at 96\,GHz,
the $J=3_{\rm k}-2_{\rm k}$ `line' is actually a blend of several
individual transitions (see Table \ref{log}) with frequencies ranging
from 145.09375\,GHz to 145.13346\,GHz, corresponding to a velocity
shift of 82\,\kms . We adopted a rest frequency of 145103.23,GHz,
corresponding to the $3_{0}-2_{0}$A$^{+}$ transition. 
The $3_{\rm k}-2_{\rm k}$ transitions were detected with relative ease
in all sources we observed, with the notable exception of M\,82. In
Maffei\,2, the $2_{\rm k}-1_{\rm k}$ line was also easily detected.
The $5_{-1}-4_{0}E$ transition in NGC\,253 is clearly detected. 

\section{Discussion}

\subsection{Physical conditions}
\subsubsection{\FORM }
Formaldehyde is a slightly asymmetric top molecule that exists in
ortho- (K$_{\rm a} = 1,3,5$\dots ) and para- (K$_{\rm a}=0,2,4$\dots )
forms that are not connected by allowed transitions. Thus, they can be
considered independent molecular species.  The asymmetry splits each
rotational level in the K$_{\rm a} > 0 $-ladders into a so-called
$K$-doublet (see, \eg , Henkel (1980) for a level diagram and more
details).

Since relative populations of different $K_{\rm a}$ ladders are
determined by collisions, line intensity ratios of transitions from
different $K_{\rm a}$ ladders of a given species are excellent tracers
of the gas kinetic temperature (Mangum \& Wootten 1993). The level
populations {\em within} a $K_{\rm a}$ ladder are determined by an
equilibrium between collisional excitation and spontaneous and
collisional deexcitation, making the relative level populations within
a $K_{\rm a}$ ladder sensitive to the H$_2$ density and, to a lesser
degree, to kinetic temperatures.

\begin{figure} 
\psfig{file=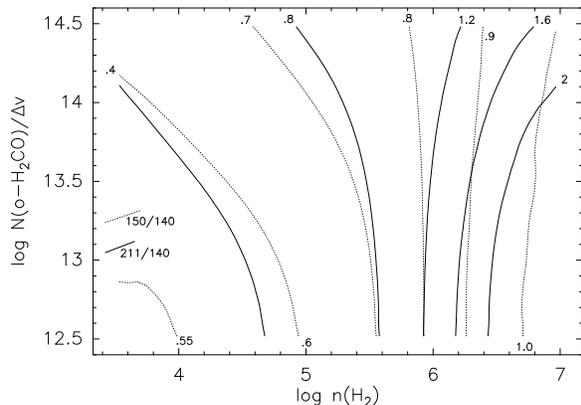,height=5.7cm,angle=-90}
\caption[]{ \label{hhcomodel} Predicted line intensity ratios of 
ortho-formaldehyde lines as a function of $n(\rm H_2)$ 
(in cm$^{-3}$) and $N$(ortho-H$_2$CO)/$\Delta v$ 
(in cm$^{-2}$/(km\,s$^{-1}$). Solid contours:
$T_{\rm B} (211{\rm \,GHz})/T_{\rm B} 140{\rm \,GHz})$, 
thin contours:  $T_{\rm B} ({\rm 150\,GHz)}/T_{\rm B}({\rm 140\,GHz})$. 
The assumed $T_{\rm kin}$ is 80\,K.}
\end{figure}

We have carried out large velocity gradient (LVG) statistical
equilibrium computations of the population of the $K_{\rm a}=1$ ladder
of ortho formaldehyde (o-\FORM ), for which we have the most complete
data.  The model is described in detail in Henkel et al.\ (1980).  We
included 16 levels of that rotational ladder fixing the kinetic
temperatures to a value of 80\,K. Setting \TKIN\ to 50\,K changes the
line intensity ratios only by a few percent. As a result of these
calculations, line ratios for the 140, 150 and 211\,GHz transitions of
o-\FORM\ are related to the H$_2$ density and the \FORM\ column density
per unit linewidth (see Fig.\,\ref{hhcomodel}).

It is evident from Fig.\,\ref{hhcomodel} that the 1.3\,mm and 2\,mm
lines have similar intensities if densities exceed $10^5\,\rm cm^{-3}$
(this also holds for para formaldehyde).  In order to derive column
densities of o-H$_2$CO from the integrated intensities of one or
several lines, an assumption must be made about the H$_2$ density and,
to a lesser degree, the kinetic temperature, both of which determine
the rotational temperature, or more generally, the partition function.
An integrated intensity of 1\,K\,kms$^{-1}$ of the 140\,GHz line
corresponds e.g.\ to $N(\rm o-H_2CO)\sim 5\,10^{13}\,\rm cm^{-2}$ if one
assumes a low H$_2$ density of 10$^4$\,cm$^{-3}$; if, however, $n(\rm
H_2)=10^{5} - 10^7 \rm \,cm^{-3}$ is assumed, then the o-H$_2$CO column
densities for the same line intensity are an order of magnitude lower.
In Table\,\ref{col}, we estimate $N(\rm o-H_2CO)$ from the intensities
of the 140\,GHz lines, and the densities from the measured line
intensity ratios.  The most reliable densities from mm-wave line
intensity ratios are estimated for regions where $n$(H$_2$) $>
10^5$\,cm$^{-3}$. For lower densities, mm-wave lines of \FORM\ become
optically thick if $N({\rm o-H_2CO})/\Delta v \ga
10^{12.5}$\,cm$^{-2}/$\kms . In this case, the line intensity ratios
not only depend on $n$(H$_2$), but also strongly on $N({\rm
o-H_2CO})/\Delta v$.  Upper limits on  $n$(H$_2$) estimated in
Table\,\ref{col} are based on the optically thin limit.

In order to determine relative abundances, the H$_2$ column density has
to be known.  In Table\,\ref{col}, we use values deduced from CO
measurements, and the standard conversion formula of Strong
\etal\ (1988), i.e.  $\eta N({\rm H}_2)/I({\rm CO}) = 2.3
10^{20}$\,\cmsq\,(\Kkms )$^{-1}$. However, there is growing evidence
that this conventional conversion factor cannot be applied to Galactic
bulge regions (see Dahmen \etal\ 1996a,b for our Galactic center,
Mauersberger \etal\ 1996a,b for NGC\,253 and NGC\,4945, Shier
\etal\ 1994 for IR-luminous galaxies). The correction factor $\eta$
might be as large as 10, and is likely to be different for different
galaxies. It will, however, always decrease H$_2$ column densities and
increase the relative abundance of the molecule.

\paragraph{NGC\,253.} Toward this galaxy, we have the largest data
base.  In order to compare line intensities which have been obtained at
different wavelengths one has to take into account the different beam
sizes and the spatial structure of the source.  Since we did not map
the distribution of H$_2$CO toward any of our sources, we assume that
the distribution of this molecule toward NGC\,253 is similar to that of
the $J=2-1$ line of $^{12}$CO, which has been mapped with 12$''$
resolution and convolved to different resolutions by Mauersberger et
al.\ (1996a). From an interpolation of the data toward the central
position of NGC\,253 in their Table\,1, intensities in a 16$''$ beam
are a factor 0.85 weaker than data at 12$''$ resolution.  In the
following analysis we assume the same correction factor for the 2\,mm
data of M\,82 and IC\,342 where the size of the molecular emission is
similar to that of NGC\,253.

We can use the intensity ratios of lines within the $K_{\rm a}=1$ ladder
of ortho formaldehyde to estimate the H$_2$ density of the H$_2$CO
emitting gas. In practice, a difficulty in assessing the line ratios
arises if the line shapes and/or the velocity coverage differs. One might 
then incorrectly compare different gas components. The $3_{13}-2_{12}$ 
(211\,GHz) and the  $2_{12}-1_{11}$ (140\,GHz) lines were observed toward
slightly different positions, adding to the uncertainty. If only the 
clear overlap region in the spectra is considered, a line intensity
ratio of 0.42($\pm$0.07) results. From Fig.\,\ref{hhcomodel} such a low 
ratio is compatible with  $n(\rm H_2)\sim 10^4\,cm^{-3}$, and even 
lower densities if H$_2$CO lines are saturated. We conclude that
$n(\rm H_2)<10^{5}\,cm^{-3}$ for the bulk of the H$_2$CO emitting gas.
From our limit to the  $3_{03}-2_{02}$ line and the detection of the
$2_{02}-1_{01}$ line of para H$_2$CO, the corresponding ratio is
$<0.2$ which is compatible with $n({\rm H_2}) \leq 10^4\,\rm
cm^{-3}$.

The ratio of the $2_{11}-1_{10}$ (ortho, 150\,GHz) and the
$2_{02}-1_{01}$ (para, 146\,GHz) transitions depends on the
ortho/para (o/p) ratio.  For the high temperature limit of the o/p
ratio of 3, one would expect a line intensity ratio of $\sim 1.4$ (for
$T_{\rm kin}=80\,\rm K$) and $n({\rm H_2})=10^4\,\rm cm^{-3}$.  The
value we actually observe is 0.6, or even lower, if one argues that
not the entire velocity range for the wide $2_{11}-1_{10}$ line should
be considered for the ratio. We therefore suggest that the o/p
ratio of H$_2$CO actually has a value close to 1, which is lower than 
the value derived by Aalto et al.\ (1997) for the 1\,mm transitions. 
Since the o/p ratio bears important information on the formation of 
molecular gas, and our result, involving a blended line, is uncertain,
further observations are very desirable.

\paragraph{M\,82.}
We have measured ortho-H$_2$CO lines toward two positions in M\,82,
namely toward the nucleus and toward the so called SW molecular hotspot
at an offset of $(\Delta \alpha, \Delta \delta)=(-10'',-10'')$
(Mauersberger \& Henkel 1991). The ratio (for the same beam size)
of the  $3_{13}-2_{12}$ (211\,GHz) and the $2_{12}-1_{11}$ (140\,GHz)
lines is $\leq 0.2$ toward the nucleus, regarding the tentatively detected
$3_{13}-2_{12}$ as contributing an upper limit only. Toward the SW
hotspot, however, it is as high as 1.2($\pm0.4$), or even $\sim 1.5$, if
we restrict the intensity from the $2_{12}-1_{11}$ line to the width of the
slightly narrower $3_{13}-2_{12}$ transition. Thus, toward the nucleus, 
$n(\rm H_2)<10^4\,\rm cm^{-3}$; while densities are as high as $n(\rm
H_2)\sim10^{6}\,\rm cm^{-3}$ toward the SW hotspot.  This estimate is
relatively robust to whether or not the H$_2$CO lines are saturated.

Such density variations were already predicted from an analysis of
centimeter wave transitions of ortho formaldehyde observed with lower
angular resolution by Baan et al. (1990).  While the 6\,cm line shows a
broad absorption (Graham et al. 1978) over the entire velocity range,
the 2\,cm line is seen in emission over a velocity range that hints
toward an origin which is confined toward the southwestern part of the
nuclear region.  The 2\,cm H$_2$CO line is usually observed in
absorption, even against the 2.7\,K cosmic background. 2\,cm emission
requires H$_2$ densities exceeding 10$^{5.5}\,\rm cm^{-3}$. This is in
very good agreement with our results from millimetric lines toward
the SW hotspot. Also the central concentration of N$_2$H$^+$, a
molecular ion that is destroyed in a high density medium,  indicates
that the gas toward the central region has a much lower density than
toward the SW hotspot (Mauersberger \& Henkel 1991).

From the observed line intensity ratio of the $2_{12}-1_{11}$  and
$2_{02}-2_{01}$ lines of $\sim 1.2$, the o/p ratio toward the SW
hotspot is (assuming $n({\rm H_2})=10^{6}\,\rm cm^{-3}$) $\sim 2$,
i.e. close to the high temperature limit. Note, however, that, as for 
NGC\,253, this result may be affected by the uncertainty of the fit to
the blended $2_{02}-2_{01}$ transition.

\paragraph{IC\,342.}
This nearby ($\sim 2$\,Mpc; McCall 1989, Karachentsev \& Tikhonov 1993)
and nearly face-on spiral galaxy is very similar to the Milky Way
galaxy with respect to the IR luminosity and the gas mass in its
central region.

We have observed formaldehyde toward the nucleus ($0'',0''$), where the
velocity coverage and the line shapes of all transition are in excellent 
agreement, and toward one offset position, $(0'',15''$), where the
(tentative) $3_{13}-2_{12}$ transition seems to appear at a different 
velocity.  The beam toward the ($0'',0''$) position also contains the
molecular clump B (Downes et al.\ 1992), which is associated with 
free-free emission equivalent to 300 O5 stars or 30 times more than the 
emission from the Galactic center star forming region Sgr\,B2. 

$3_{13}-2_{12}$/$2_{11}-1_{10}$ (211\,GHz/150\,GHz) line ratios indicate 
high densities of $10^{6.2} \rm cm^{-3}$ toward the $(0'',0'')$ position.
The limit to the $2_{11}-1_{10}$/$2_{12}-1_{11}$ (140\,GHz/150\,GHz)
line ratio toward the nucleus of $\geq 1$ is also suggesting  
high densities, while the limit toward the ($0'',15''$) offset ($\leq 0.4$, 
considering the tentative detection to be an upper limit, and regarding a
common velocity interval for both lines), indicates a far lower density 
of $< 10^4$\,\percc . As in M\,82, our two beams pick up contributions 
from two molecular phases, namely from a moderately dense interclump gas 
and from the clumps themselves.  The presence of several molecular gas
components has also been inferred by Downes et al.\ (1992) from the
comparison of single dish and interferometric observations in CO.

\paragraph{Maffei\,2.} 
The ratios of the 150 and 140\,GHz line, which agree very well in
velocity coverage and line shape, indicate typical densities of the order 
of $10^4\,\rm cm^{-3}$.

\subsubsection{\METH }

Methanol is an asymmetric top molecule capable of hindered internal
rotation. Transitions from A-type (internal rotation) to E-type (no
internal rotation) levels are strictly forbidden; the two types behave
as separate molecular species in the ISM. In addition, $a$ and $b$-type
transitions have to be distinguished according the component of the
dipole moment that is is parallel to the axis of rotation.  All
transitions contributing to the $3_{\rm k}-2_{\rm k}$ and $2_{\rm
k}-1_{\rm k}$ lines are $a$-type ($\Delta k=0; \mu_a=0.896$\,Debye),
while the $5_{-1}-4_{0}$ transition is $b$-type ($\Delta k=1;
\mu_b=1.412$\,Debye) (Lees et al.\ 1973, Sastry et al.\ 1981; for a
detailed discussion of the \METH\ molecule, see \eg\ Menten 1987).

\begin{figure}
\psfig{file=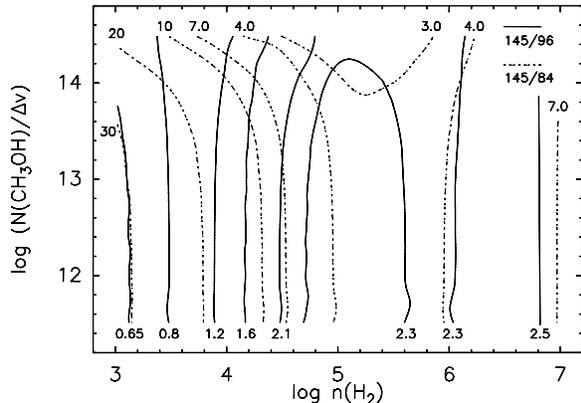,height=5.7cm,angle=-90}
\caption[]{ \label{methLVG} Predicted intensity ratios of methanol 
lines for a raster of values for $n(\rm H_2)$ (in cm$^{-3}$) and 
$N{\rm (CH_3OH)})/\Delta v$ (in cm$^{-2}$/(km\,s$^{-1}$). Solid 
contours: $T_{\rm B} ({145\rm \,GHz})/ T_{\rm B} (96{\rm \,GHz})$, 
dashed contours:  $T_{\rm B} ({\rm 145\,GHz)}/T_{\rm B}({\rm 84\,GHz})$. 
The assumed $T_{\rm kin}$ is 80\,K.}
\end{figure}

An analysis of the $3_{\rm k}-2_{\rm k}$ group of transitions is
difficult since the lines are blended.  We have, however, carried out
LVG calculations analogous to those for \FORM, using a program provided
by C.M.\ Walmsley (see, e.g.\ Walmsley \etal\ 1988, Bachiller
\etal\ 1995).  We have modelled the intensity ratios of the sum of all
transitions contributing to the $3_{\rm k}-2_{\rm k}$ group to the sum
of the $2_{\rm k}-1_{\rm k}$ transitions and the ratio of the $3_{\rm
k}-2_{\rm k}$ group to the $5_{-1}-4_{0}E$ line as a function of H$_2$
density and \METH\ column density per unit linewidth
(Fig.\ \ref{methLVG}), for a kinetic temperature of 80\,K.  Clearly,
for $N$(\METH )$/\Delta v \la 10^{14}$\,cm$^{-2}$/km\,s$^{-1}$, the
line ratios are independent of column density. Thus, the \METH\ lines
are very likely to be optically thin in all our sources.  We have used
the H$_2$ densities determined from Fig.\ \ref{methLVG} and the
integrated intensities of the $3_{\rm k}-2_{\rm k}$ group of
transitions to determine the \METH\ column densities given in
Table\,\ref{col}.

\begin{table}
\begin{flushleft}
\caption{ \label{col} Beam averaged \FORM\ and \METH\ column densities 
and relative abundances. }
\begin{tabular}{l@{}|c@{}c@{}c@{}r@{}r}
\multicolumn{1}{c}{Source} & \multicolumn{1}{c}{$\eta N$} & 
\multicolumn{1}{c}{$N$} & \multicolumn{1}{c}{$N$} &
\multicolumn{1}{c}{$\frac{[\FORM ]}{[\MOLH ] \eta}$} &
\multicolumn{1}{c}{$\frac{[\METH ]}{[\MOLH ] \eta}$} \\
 & \multicolumn{1}{c}{H$_2$$^{a)}$} & \multicolumn{1}{c}{\FORM } & 
 \multicolumn{1}{c}{\METH } & &  \\
 & \multicolumn{1}{c}{$10^{22}$}  & \multicolumn{1}{c}{$10^{13}$}  
& \multicolumn{1}{c}{$10^{13}$} & \multicolumn{1}{c}{$10^{-9}$} &   
\multicolumn{1}{c}{$10^{-9}$} \\
 & \multicolumn{3}{c}{\cmsq }  &  & \\
\hline
NGC\,253   & & & & & \\
(0,0)      & 37  & 40$^{b)}$ & 50$^{b)}$  & 1.1  & 1.4 \\         
M\,82      & & & & & \\ 
(0,0)      & 16  & $\sim 25^{c)}$ & $\leq$\,2$^{e)}$ & 1.6& $\leq 0.1$ \\
(--10,--10) &    & 0.9$^{d)}$  & $\leq$\,1--2$^{e)}$  & & \\
IC\,342    & & & & & \\
(0,0)      & 4.9 & 0.2$^{d)}$ & & 0.04  & \\
(0,5)      & 4.9 & & 11$^{f)}$   & & 2.2 \\
(0,--5)    &     & & 6$^{f)}$        & & \\
(0,15)     &     & $>8^{c)}$  &  6$^{f)}$         & & \\
Maffei\,2  & 1.4 & 15$^{b)}$  &  4$^{b)}$ & 11 &  2.9 \\
NGC\,6946  & 2.3 & &  3$^{b)}$       & & 1.3 \\
\hline
\end{tabular} \\
$a)$ based on $I$(CO) (see H\"uttemeister \etal\ 1995 and references 
therein) and a ``conventional'' conversion factor. \\
$b)$ assumed $n({\rm H}_2)$: $\sim 10^4$\,\percc \\
$c)$ assumed $n({\rm H}_2)$: $< 10^{4}$\,\percc \\
$d)$ assumed $n({\rm H}_2)$: $\sim 10^{6}$\,\percc \\
$e)$ from Eq.\,1; with \TROT\,=10\,K \\
$f)$ assumed $n({\rm H}_2)$: $\ga 10^3$\,\percc   \\
\end{flushleft}
\end{table}

\paragraph{NGC\,253.} The map we obtained in the $3_{\rm k}-2_{\rm k}$
transition toward NGC\,253 (Fig.\ \ref{methf2}) shows the same general
shape as maps in other high density tracing molecules like HNC
(H\"uttemeister et al.\ 1995) and CS (Mauersberger \& Henkel 1989). The
emission extends from the north-east to the south-west, with a (not
deconvolved) full source size of $\sim 55'' \times 25''$, which is
similar to the extent of intense $^{12}$CO(2--1) emission (Mauersberger
et al.\ 1996a).
 
Therefore, it seems reasonable to correct the $3_{\rm k}-2_{\rm
k}$/$2_{\rm k}-1_{\rm k}$ and the $3_{\rm k}-2_{\rm k}$/$5_{-1}-4_{0}E$
line ratios for the different beamsizes in the same way as for
\FORM\ (Section 4.1.1), based on the Mauersberger et al.\ (1996a) data.
The line intensities of the 3\,mm lines then have to be multiplied by
$\sim 1.4$ to make them directly comparable to the 2\,mm lines. The 
linewidths of the two groups of transitions agree well. The resulting
$3_{\rm k}-2_{\rm k}$/$2_{\rm k}-1_{\rm k}$ ratio (0.8$\pm$0.15)
implies a H$_2$ density of $\sim 10^{3.5}$\,\percc. The $5_{-1}-4_0$
line is much narrower than the $3_{\rm k}-2_{\rm k}$ group. Since we 
are comparing a group of transitions spaced by $> 80$\,\kms\ to an 
unblended line, this is to be expected. However, it might not account for 
the entire difference. The ratio of the integrated intensities 
(6.8$\pm$2.4) thus is an upper limit, yielding a lower limit to the density of 
$10^{4.5}$\,\percc . If we consider only the velocity interval for the 
$3_{\rm k}-2_{\rm k}$ group where $5_{-1}-4_0$ emission is present,
this lower limit to the intensity ratio of 3.1$\pm$0.7 corresponds to a 
an upper limit to the H$_2$ density of $\geq 10^5$\,\percc (see Fig.\ 
\ref{methLVG}). In any case, the values from the 
$3_{\rm k}-2_{\rm k}$/$2_{\rm k}-1_{\rm k}$ and the 
$3_{\rm k}-2_{\rm k}$/$5_{-1}-4_0$ line ratios embrace the H$_2$ density
estimated from \FORM .  The critical density of the $5_{-1}-4_{0}E$
transition is smaller than the mean of the $3_{\rm k}-2_{\rm k}$ group
by a factor of $\sim 5$, while the difference is much less pronounced
between the $3_{\rm k}-2_{\rm k}$ and the $2_{\rm k}-1_{\rm k}$
groups.  This might explain the difference in the density estimates
resulting from the two line intensity ratios.

At an H$_2$ density close to $10^{4}$\,\percc, both the $2_{\rm
k}-1_{\rm k}$ and the $3_{\rm k}-2_{\rm k}$ groups of lines are
subthermally excited (\TEX $<$ \TKIN), with excitation temperatures for
the individual transitions ranging  from 10\,K to 20\,K.  We thus
obtain a value for $N$(\METH ) which is lower by a factor of $\sim 6$
than what is determined under the assumption that the density is
sufficiently high to thermalize the \METH\ emission (Henkel \etal\
1987).

\paragraph{M\,82.} The clear non-detection of any methanol line in
M\,82 is one of our most striking results. Henkel \etal\ (1987)
searched for the $2_{\rm k}-1_{\rm k}$ transition in this galaxy and
established an upper limit of 30\,mK rms \TMB\ for this line.  We obtained
upper limits of 10--15\,mK rms for the $3_{\rm k}-2_{\rm k}$
transition toward the central position and the north-eastern and
south-western peak positions of the molecular ring. 

Since no line ratios could be determined, we have used the LTE
approximation for optically thin lines to estimate the total
\METH\ column density:
\begin{equation}
\label{eq1}
N(\METH ) = 1.28 \, T_{\rm rot}^{1.5} \, e^{E_{\rm l}/k T_{\rm rot}} 
\frac{1.67 \, 10^{14}}{\nu \mu^2 S} \int T_{\rm MB} dv
\end{equation}
($E_{\rm l}$: Energy of lower level, $\nu$: line frequency in GHz,
$\mu$: dipole moment in Debye, $S$: linestrength; Menten et
al.\ 1988). 

Using Eq.\,\ref{eq1}, we have fitted all components of the 
$3_{\rm k}-2_{\rm k}$ group to the limit to the integrated intensity.
For subthermally excited lines (\TROT = 10\,K), we find column
densities ranging from smaller than 1\,$10^{13}$\,\cmsq\ (SW hotspot)
to less than 2\,$10^{13}$\,\cmsq\ (center and NE hotspot),
corresponding to an abundance of [\METH ]/$\eta$[\MOLH ]$\leq 1.2
10^{-10}$ (see Table \ref{col}). If we assume that the H$_2$ densities
are the same as the ones found for \FORM\ and apply our LVG model, the
only change is a slight increase in the limit for $N$(\METH ) to
2\,$10^{13}$\,\cmsq\ in the dense SW hotspot (see Table\,\ref{col}). 

\paragraph{IC 342.} Within the $3_{\rm k}-2_{\rm k}$ blend, two groups
of lines are separated by $\sim 45$\,\kms. The group of 5 lines at
higher frequency  has generally higher energies than the lower
frequency group. Because the lines in IC\,342 are narrow, these two
groups are resolved, and the higher energy transitions are below the
detection limit. Thus, LTE fits to the $3_{\rm k}-2_{\rm k}$ group
already greatly constrain the possible excitation conditions:
\TROT\ has to be 10\,K or lower.  Since the kinetic temperature of the
gas in IC\,342 is likely to be at least 50\,K (Ho et al.\ 1990),
\METH\ has to be very subthermally excited, with \TEX \ lower than for
NGC\,253.

The LVG calculations confirm this: For a (beam size corrected) $3_{\rm
k}-2_{\rm k}$/$2_{\rm k}-1_{\rm k}$ intensity ratio of 0.6($\pm$0.1),
observed toward the (0$'',$5$''$) position, we find an H$_2$ density of
only $\sim 10^{3}$\,\percc\ (\TKIN = 80\,K) to $\sim
10^{3.4}$\,\percc\ (\TKIN = 50\,K) and excitation temperatures of $\sim
5 - 10 $\,K. This implies that \METH\ is sensitive to interclump gas 
of only low to moderate density.

The beam-averaged column density we derive is
1.1\,$10^{14}$\,\cmsq\ for a position slightly north of the center of
IC\,342, and half that for offsets to the south and further to the
north (assuming the same low density), corresponding to an abundance of
[\METH ]/$\eta$[\MOLH ]$\,\sim 2\,10^{-9}$ (Table \ref{col}).  We note
that the column densities given in Henkel \etal\ (1988), based on
\TROT\ = 50\,K and derived from the $2_{\rm k}-1_{\rm k}$ line, are far
too high. A recalculation using \TROT\ = 8\,K and Eq.\,\ref{eq1} (with
$\sum S_i = 7$ and $E_l =6$\,K) yields a total column density that
agrees to within 10\% with our result for the $3_{\rm k}-2_{\rm k}$
line, showing that the LTE and LVG approximations converge in this
case.  The limit given by the non-detection of the $5_{-1}-4_{0}$
transition is also compatible with a total \METH\ column density of
$\sim 10^{14}$\,\cmsq.

\paragraph{Maffei\,2 and NGC\,6946.} 
The $3_{\rm k}-2_{\rm k}$/$2_{\rm k}-1_{\rm k}$ line ratios, corrected
for beam size, imply H$_2$ densities of $\sim 10^4$\,\percc . It thus
seems likely that the excitation of \METH\ in both galaxies is
similar to what is found in NGC\,253.

\subsection{Chemical variations}

\paragraph{Molecular abundances.} For most of the galaxies observed,
the column densities of o-H$_2$CO and CH$_3$OH are comparable. This
might also be the case for the M\,82 SW hotspot, where we have only a
limit for the CH$_3$OH column density. The notable exception is the
nucleus of M\,82 where the column density of CH$_3$OH is at least an
order of magnitude lower than that of H$_2$CO. It is quite clear that
this is not just mimicked by excitation effects.  Such a discrepancy
between M\,82 and other starburst galaxies, especially NGC\,253, 
has already been noticed before (e.g.\ Mauersberger \& Henkel 1993).

Evaporation from grain surfaces is in many cases a crucial process to
maintain a high gas phase abundance of complex molecular species. Models 
predict that these molecules, such as CH$_3$OH and also \FORM , are 
chemically converted into simpler species on a  timescale of order 10$^5$ 
years (Helmich 1996).  Methanol evaporates at a temperature around 70\,K
(Turner 1989, Nakagawa 1990)  and is known to be a tracer of hot, dense
gas (Menten et al.\ 1988), since its abundance is observed to increase
dramatically in the vicinity of young massive stars.

Formaldehyde, on the other hand, empirically shows far smaller
abundance variations in Galactic interstellar clouds (Mangum and
Wootten 1993) although it is 3--25 times more abundant in the Orion hot
core than in other molecular clouds (Mangum et al.\ 1990). Contrary to
CH$_3$OH, H$_2$CO is also abundant in cirrus clouds (Turner 1993) and
cool, dense Galactic disk sources. It seems certain that processing on
dust grains must play a role in the chemistry of \FORM , but the
replenishment process is unclear (Federman \& Allen 1991, Turner 1993,
Liszt \& Lucas 1995).

Takano et al. (1995) notice that all molecules that are known to be 
depleted in M\,82 (besides CH$_3$OH also SiO, HNCO, CH$_3$CN and SO) 
form preferentially under high temperature conditions.
In the presence of a steep gravitational potential and a central bar,
we expect shocks and turbulence to play a significant role in the
heating of molecular clouds distant from the actual sites of star
formation (e.g.\ H\"uttemeister et al.\ 1993, Das \& Jog 1995). This
can explain the high temperatures toward the central molecular
condensation of NGC\,253, which is much more compact than that of
M\,82, a smaller, irregular galaxy. Tidal heating should therefore
operate less efficiently in M\,82 than in larger spirals like NGC\,253,
NGC\,6946 or IC\,342.  Thus, one would expect the bulk of the gas
in the center of NGC\,253 to be warmer than that toward the center of
M\,82 (with the exception of those clouds which are heated directly by
newly formed stars).

The abundance of \METH\ in NGC\,253 does not seem to be exceptionally
high when compared to NGC\,6946 or non-starburst nuclei like Maffei\,2
and IC\,342.  But we tentatively find that the gas traced by \METH \ in
NGC\,253 is at a similar density as the gas traced by \FORM, while the
gas traced by \METH\ close to the nucleus of IC\,342 is at a lower
density. If temperature is indeed the key to the distribution of \METH ,
we may conclude that the warm gas in IC\,342 is less dense than in
NGC\,253.  Since IC\,342 is not a starburst galaxy, this agrees with
the fact that warm gas in the center of our Galaxy is, in the absence
of massive star formation, at lower densities than cooler gas
(H\"uttemeister et al.\ 1993), while the opposite is expected for
active star forming regions. NGC\,253 might be lacking the high
temperature, low density interclump gas component picked up by
\METH\ in IC342; the bulk of the molecular material in NGC\,253 is 
warm, at least moderately dense gas.  In this scheme, M\,82 has a cooler 
interclump component, seen in \FORM\ and  N$_2$H$^+$ (Mauersberger \& 
Henkel 1991), but not in \METH .

\paragraph{Extended methanol emission in NGC\,253.} While the general
shape of the central molecular condensation seen in \METH\ and other
molecules is similar, in a detailed comparison differences are
apparent.  The maps of total integrated intensity in $^{12}$CO(2--1),
CS and HNC all show only one central peak. Only when the red- and
blueshifted emission is displayed separately, two distinct peaks,
separated by $10'' - 15''$, become visible, with the redshifted
emission centered toward the south-west. The \METH\ map
(Fig.\ \ref{methf2}) shows these two peaks even for the total velocity
range. The positions of the peaks agree to within better than $5''$
with the `red' and `blue' peaks seen in the other molecules. From the
channel maps, the blueshifted emission is strongest at 200\,\kms\ and
the red-shifted emission peaks at 300\,\kms. In addition, there is a
third peak, visible in the intensity and channel maps, at an offset of
$\sim$(--25$''$,--10$''$), at an even higher velocity of 350 --
450\,\kms.

Only the interferometric $^{12}$CO(1--0) map obtained by Canzian et
al.\ (1988) with a $5'' \times 9''$ beam also shows two resolved peaks
(at the same positions as the \METH\ peaks) when the total velocity
range is considered.  Thus, the single dish \METH\ data match the
interferometric data not only more closely than the single dish
$^{12}$CO data, but also better than single dish maps of other high
density tracers, namely CS and HNC.  Since the interferometer misses
extended emission, we can conclude that \METH\ traces a more confined
gas component than CS or HNC. Since the critical densities of
\METH($3_{\rm k}-2_{\rm k}$), CS(2--1) and HNC(1--0) are roughly
similar at $\sim 3 - 6 \, 10^5$\,\percc , the differences in spatial
distribution seem to be caused by chemical fractionation rather than
gas density. Since production of \METH \ is favored in high
temperatures, the differences in distribution may reflect differences
in gas temperatures within the generally at least moderately dense gas 
in the bulge of NGC\,253, with \METH\ preferentially tracing the sites
of ongoing massive star formation.

\section{Conclusions}

We have observed rotational transitions of the high gas density tracing
molecules \FORM\ and \METH\ toward a sample of nearby gas rich external
galaxies, namely NGC\,253, Maffei\,2, IC\,342, M\,82 and NGC\,6946. Our
main results are:

\begin{enumerate}

\item H$_2$CO was detected in all 5 galaxies searched (tentatively in
NGC\,6946).

\item In the prominent starburst galaxy NGC\,253, the bulk of the gas
emitting \FORM\ lines is at densities of $\sim 10^4$\,\percc .  In both
M\,82 and IC\,342, the \FORM\ line ratios distinguish two different
components of molecular gas. High densities dominate the SW molecular
hotspot in M\,82 and the central region of IC\,342 ($n({\rm H_2})
\approx 10^{6}$\,\percc ), while lower densities  ($< 10^4$\,\percc)
are prevalent toward the center of M\,82 and the north of the nucleus
of IC\,342.

\item We detected \METH\ lines in the same sample of galaxies, with the
notable exception of M\,82. 

\item In all galaxies detected, methanol is subthermally excited. In
NGC\,253, Maffei\,2 and NGC\,6946, moderate gas densities of $\sim
10^4$\,\percc\ are likely. Toward a position north of the center of
IC\,342, the H$_2$ densities seem to be  at $\sim
10^{3}$\,\percc. This might indicate that warmer gas is at lower
densities in the non-starburst nucleus of IC\,342, where  
\METH\ traces an entirely different gas component than \FORM .

\item In M\,82, no methanol could be detected down to an rms level of
10--15\,mK, corresponding to an abundance at least an order of
magnitude lower than in the other galaxies. This confirms large scale
chemical differences in starburst centers. The likeliest cause of these
are global differences in the temperature of the molecular material.

\item A map of NGC\,253 in the $3_{\rm k}-2_{\rm k}$ line of methanol
shows several peaks of emission and a greater amount of spatial
structure than than single-dish maps of of other molecules, even
high-density tracers.  Thus, \METH\ emission is associated with smaller
scale clumps.  This indicates chemical fractionation not only between
different galaxies but also within the bulge of NGC\,253, possibly
reflecting the temperature structure of the gas.

\end{enumerate}

\acknowledgements{R.M. was supported by a Heisenberg fellowship by the
Deutsche Forschungsgemeinschaft. We thank K.\ Menten and C.M.\ Walmsley
for help with the interpretation of the methanol data,
S.\ Aalto-Bergman, P.\ Jewell, J.\ Mangum, and S.\ Radford for useful
discussions, and the referee, L.-\AA.\ Nyman, for helpful comments. }

\end{document}